\documentclass[%
 aip,
 amsmath,amssymb,
 reprint,%
]{revtex4-1}

\usepackage{graphicx}
\usepackage{dcolumn}
\usepackage{bm}

\usepackage[utf8]{inputenc}
\usepackage[T1]{fontenc}
\usepackage{mathptmx}
\usepackage{etoolbox}
\usepackage{hyperref}

\makeatletter
\def\@email#1#2{%
 \endgroup
 \patchcmd{\titleblock@produce}
  {\frontmatter@RRAPformat}
  {\frontmatter@RRAPformat{\produce@RRAP{*#1\href{mailto:#2}{#2}}}\frontmatter@RRAPformat}
  {}{}
}%
\makeatother

\usepackage{xcolor}
\newcounter{aqctr}
\newenvironment{author-query}
{\refstepcounter{aqctr}\par\vspace{\baselineskip}\noindent
\color{red}\textbf{Author Query/Comment AQ \arabic{aqctr}.}}
{\par\vspace{\baselineskip}\normalcolor}

\begin{document}

\preprint{AIP/123-QED}

\title{Utilizing the Janus MoSSe surface polarization in designing complementary metal-oxide-semiconductor field-effect transistors}
\author{Yun-Pin Chiu}
\author{Hsin-Wen Huang}%
\author{Yuh-Renn Wu}
 \email[Author to whom correspondence should be addressed: ]{yrwu@ntu.edu.tw}
\noaffiliation
\affiliation{ Graduate Institute of Photonics and Optoelectronics and Department of Electrical Engineering, National Taiwan University, Taipei 10617, Taiwan
}%

\date{\today}

\begin{abstract}
Janus transition metal dichalcogenides (JTMDs) have attracted much attention because of their outstanding electronic and optical properties. The additional out-of-plane dipole in JTMDs can form n- and p-like contacts, and this may be used in device applications, such as pin diodes and photovoltaic cells. In this study, we exploit this property to design n- and p-type metal-oxide-semiconductor field-effect transistors (MOSFETs). First, we use density-functional-theory calculations to study the inherent dipole field strength in the trilayer JTMD MoSSe. The intrinsic dipole of MoSSe causes band bending at both the metal/SMoSe and SMoSe/metal interfaces, resulting in electron and hole accumulation to form p-type, n-type, and even  Ohmic contact regions. We incorporate this property into a two-dimensional finite-element-based Poisson-drift-diffusion solver to perform simulations, based on which we design complementary MOSFETs. Our results demonstrate that JTMDs can be used to make n- and p-MOSFETs in the same layer without the need for any extra doping. 
\end{abstract}

\maketitle

%

\section{Introduction}

In the field of two-dimensional (2D) semiconducting materials, research on transition metal dichalcogenides (TMDs) has been flourishing in recent years owing to their remarkable properties, such as relatively high carrier mobilities,\cite{2D_TMD_intro_mobility1,2D_TMD_intro_mobility2,2D_TMD_intro_mobility3} suitable band-gap energies,\cite{2D_TMD_intro_bandgap1,2D_TMD_intro_bandgap2} and an atomic thickness that is modifiable.\cite{2D_TMD_intro_thickness1,2D_TMD_intro_thickness2,2D_TMD_intro_thickness3} These properties make 2D TMDs suitable for various device applications\cite{2D_transistors1, TMD_CMOSFET_Chart, Ying-Chuan_KMC, CFET_IEEE,2D_Photo}, especially for replacing traditional channel materials and overcoming the barrier of short-channel effects in smaller field-effect transistors (FETs).  \cite{2D_transistors1,2D_TMD_intro_mobility1,2D_FET_Pin-Fang} These TMD-based channel FETs have been studied for over a decade, and FETs have been successfully fabricated with TMD materials. For example, MoS$_{2}$ and WS$_{2}$ are  two commonly selected materials for n-type channels, and $\alpha$-MoTe$_{2}$ and WSe$_{2}$ are both good candidates for p-type channels.\cite{2D_TMD_intro_mobility3,TMD_CMOSFET_1,TMD_WS2_FET,TMD_CMOSFET_Chart,TMD_WSe2_FET}

It is well known that the contact characteristics of a FET device will strongly affect its $I$-$V$ characteristics and performance. For low contact resistance, Ohmic contacts are usually required at the semiconductor-metal interface to enhance the electrical performance of FETs while maintaining a linear $I$-$V$ curve.\cite{sze2021physics, TMD_contact_1, TMD_contact_5} Unfortunately, Schottky barriers (SBs) are usually formed for most   TMD/metal contacts.\cite{TMD_contact_1, TMD_contact_2, TMD_contact_3, TMD_contact_4, TMD_contact_5} For a FET with SB contacts, carriers must overcome the energy barrier to transport from metal into the TMD channel region, impairing the device function by lowering the injection efficiency.\cite{SBH_1} A typical technique used to reduce contact resistance in modern semiconductor technology is substitutional doping.\cite{TMD_FET_Doping} However, there are better ways to apply such a doping method to 2D materials since their characteristics will be heavily altered after doping. Additionally, the absence of dangling bonds on the surface of a 2D material makes solid interface connections with metal challenging, resulting in larger contact resistance.\cite{TMD_contact_1} Hence, most of the effort required when engineering TMD/metal contacts is associated with finding a suitable metal to adjust the work function. For n-FETs, a material with a smaller work function is needed, whereas a metal with a high work function is often selected for p-FETs. Several studies have shown that TMD/metal heterostructures, such as MoS$_{2}$/Cu, WS$_{2}$/Cu, and WSe$_{2}$/Ti could provide n-type SB  contacts,\cite{TMD_contact_2, TMD_contact_3} and p-type SB contacts can be formed with a few combinations, such as WSe$_{2}$/Pd and WSe$_{2}$/Au.\cite{TMD_contact_2,TMD_contact_WSe2_p} However, only a specific TMD/metal combination is acceptable for Ohmic contact formation, making  TMD-based FET contact design more challenging.\cite{JTMD_Metal_Hetero} 

To combine n- and p-type FETs into a complementary metal-oxide-semiconductor FET (CMOSFET) device, horizontally combing two different TMD channels to form a CMOSFET is a common approach.\cite{TMD_CMOSFET_exp1, TMD_CMOSFET_exp2, TMD_CMOSFET_Chart, CFET_IEEE} With regard to fabrication, however, this CMOSFET design with two different materials has the disadvantage that the two materials need to be transferred onto a single substrate, with the risk of introducing defects or contamination.\cite{Transfer_dis} Furthermore, two contact metals need to be considered separately, resulting in higher processing complexity and a more time-consuming fabrication procedure. Recently, ambipolar TMDs, such as MoTe$_{2}$\cite{TMD_CMOSFET_exp3} and ReSe$_{2}$\cite{ambipolar_TMD}, have been proposed to compose a single-channel material CMOSFET, which reduces the difficulty of device fabrication.  

The use of an alternative type of materials, namely, Janus transition metal dichalcogenides (JTMDs), MXY, where M $=$ transition metal and X, Y $=$ S, Se, or Te, with X $\neq$ Y, is another possible way to fabricate a CMOSFET using a single material. In recent years, several JTMDs have been successfully synthesized via different techniques. Monolayer MoSSe, for instance, was synthesized utilizing chemical vapor deposition combined with plasma stripping\cite{MoSSe_fabricate} or a selective epitaxy atomic replacement process,\cite{MoSSe_fabricate_2020} while Lin \emph{et al.} \cite{WSSe_fabricate_2020} prepared Janus WSSe by pulsed-laser deposition. Previous theoretical studies of these JTMDs have provided unmistakable proof that their phonon spectra are stable and devoid of imaginary frequencies, verifying their fabrication stability.\cite{DFT_JTMD_phonon, DFT_WSSe} Compared with TMDs, JTMDs have two different chalcogen-atom layers, which create an asymmetric sandwich configuration that provides an out-of-plane electric field. This out-of-plane field has the characteristic of a dipole, which enables potential applications in piezoelectric, spintronics, catalysts, and gas-sensing devices.\cite{JTMD_overview, JTMD_overview_chem, janus_gan} Also, simulation studies have shown that this intrinsic dipole feature of JTMDs can enable the formation of both n- and p-type contacts simultaneously without the need for additional doping when different sides are attached to graphene (Gr),\cite{JTMD_Graphene_Hetero, JTMD_Graphene_pin, JTMD_Graphene_photovotalic} and this can be incorporated into the design of pin-diode\cite{JTMD_Graphene_pin} or photovoltaic\cite{JTMD_Graphene_photovotalic} devices. Moreover, theoretical studies have shown that the SB at contact could be substantially reduced when JTMD MoSSe are attached to metals,\cite{JTMD_Metal_Hetero} owing to the intrinsic dipole. The above-mentioned JTMD/electrode contact systems may even attain Ohmic contact under certain conditions.\cite{JTMD_Graphene_Hetero, JTMD_Metal_Hetero} However, unlike optoelectronic devices, the exploitation of the advantages of n- and p-type Ohmic contacts in FET applications has not yet been explored. Furthermore, there has not been much work on utilizing this intrinsic dipole property to form FETs or CMOSFETs without doping. 

In this study, we propose a CMOSFET design concept utilizing the intrinsic dipole of a JTMD. The material MoSSe was selected for investigation since it was the first JTMD and is the most commonly synthesized.\cite{MoSSe_fabricate, MoSSe_fabricate_2020} The remainder of this paper is organized as follows. First, we illustrate the JTMD dipole and contact characteristics for three heterostructures,  namely, Gr/trilayer AA$'$A stacking MoSSe/Gr and A$'$A stacking bilayer MoSSe(2L-MoSSe)/Al (111) face and 2L-MoSSe/Cu (111) face structures, via first-principles density-functional theory (DFT). Then, we demonstrate the JTMD CMOSFET device design concept with simulation results using a 2D finite-element-based Poisson-drift-diffusion solver (2D DDCC) developed in our laboratory.\cite{wuoptoelectronic}

\section{Computational details}

\subsection{Heterostructure and layered MoSSe DFT simulations}
The heterostructure calculations are performed with Quantum ESPRESSO,\cite{QUANTUM} an open-source DFT calculation software package employing the generalized gradient approximation (GGA) of the exchange-correlation functional and norm-conserving  Vanderbilt\cite{NC_pp} pseudopotentials combined with DFT-D3 van der Waals correction.\cite{DFT_D3} An $8\times8\times1$ Monkhorst-Pack $k$ sampling is used. The geometry is refined until the maximal residual forces are less than 0.015~eV/\AA. At this point, MoSSe heterostructures are subjected to dipole correction to negate the artificial electric field caused by their asymmetric structures. The vacuum region is set to 20~\AA\ to prevent interactions between nearby supercells. 
For MoSSe contact simulation, we attach a $\sqrt7\times\sqrt7\times$1 supercell of Gr to a $2\times2\times1$ 2L-MoSSe supercell. A $2\times2\times1$ supercell of six layers of (111) Al to a $\sqrt3\times\sqrt3\times$1 2L-MoSSe supercell, and a $\sqrt7\times\sqrt7\times$1 supercell of six layers of (111) Cu to a $2\times2\times1$ 2L-MoSSe supercell. A simulation model is created for each metal's energetically advantageous surface,\cite{skriver1992surface}, and four outward metal layers are fixed in the relaxation process. 
For layered MoSSe, we construct an AA$'$A stacking of the MoSSe trilayer structure, with the above calculation strategy also being carried out here. The only difference is that the $k$ mesh is set to $36\times36\times6$ in the MoSSe band-structure calculation.

\subsection{Device simulation}
The 2D-DDCC solver is used to examine the device's performance. This solver was developed in our laboratory and has been used on many different kinds of devices. More detailed information can be found on our website.\cite{wuoptoelectronic}   Both the Poisson and drift-diffusion equations are solved self-consistently by the solver. The equations are as follows:
\begin{gather}
 \nabla_r\cdot[\epsilon\nabla_rV(r)]=q(N_A^--N_D^++n-p), 
 \label{poisson}
\\
  J_n=-q\mu_nn(r)\nabla_rV(r)+qD_n\nabla_rn(r),
 \label{Jn}
\\
   J_p=-q\mu_pp(r)\nabla_rV(r)-qD_p\nabla_rp(r),
 \label{Jp}
 \end{gather}
where $\epsilon$ is the dielectric constant, $V$ is the electric potential, $N^{-}_{A}$ is the acceptor's doping concentration, $N^{+}_{D}$ is the donor's doping concentration, $n$ is the electron concentration, $p$ is the hole concentration, $J_n$ and $J_p$ are the current densities of electrons and holes, respectively, and $D_n$ and $D_p$ are the electron and hole diffusion coefficients, respectively. This solver serves as the basis for all of the device simulations discussed in this paper, except the MoSSe channel region calculations.

\begin{table}[!t]
\caption{\label{parameters}. Input parameters used in the JTMD FET simulations. The parameters in the upper part of the table are  MoSSe-related constants, and those in the lower part are parameters of the oxides.}
\begin{ruledtabular}
\begin{tabular}{c c l}
Parameter&Numerical value&Description\footnote{MoSSe parameters are all obtained by DFT. SiO$_2$ and Al$_2$O$_3$ parameters are given inherently in 2D DDCC.\cite{wuoptoelectronic}}\\
\hline
$E_g$ & 1.212~eV & Average band gap\\
$E_{ea}$ & 3.97~eV & Electron affinity\\
$\mu_e$ & 667~cm$^{2}$/(V$\cdot$s) & Electron mobility\\
$\mu_h$ & 624~cm$^{2}$/(V$\cdot$s) & Hole mobility\\
$m^*_{e,K}$ & $0.37m_0$ & Electron $K$-valley effective mass\\
$m^*_{e,Q}$ & $0.47 m_0$ & Electron $Q$-valley effective mass\\
$m^*_{h,\Gamma}$ & $0.75 m_0$ & Hole $\Gamma$-valley effective mass\\
$m^*_{h,K}$ & $0.45 m_0$ & Hole $K$-valley effective mass\\
$\epsilon_{||}$ & $15.93 \epsilon_0$ & In-plane dielectric constant\\
$\epsilon_{\bot}$ & $6.31 \epsilon_0$ & Out-of-plane dielectric constant\\
$P_\mathrm{sp}$ & $4.76\times10^{13}$~cm$^{-2}$ & Spontaneous polarization\\
\hline
$\epsilon_\mathrm{SiO_2}$ & $3.9 \epsilon_0$ & SiO$_2$ dielectric constant\\
$\epsilon_\mathrm{Al_2O_3}$ & $9.3 \epsilon_0$ & Al$_2$O$_3$ dielectric constant\\
$E_{g,\mathrm{HfO}_2}$ & 5.8~eV & HfO$_2$ band gap\cite{HfO2_Para}\\
$\epsilon_\mathrm{HfO_2}$ & $25 \epsilon_0$ & HfO$_2$ dielectric constant\cite{Oxide_Para}\\
\end{tabular}
\end{ruledtabular}
\end{table}

\section{Results and discussion}

\begin{figure}[!t]
\includegraphics[width=\columnwidth]{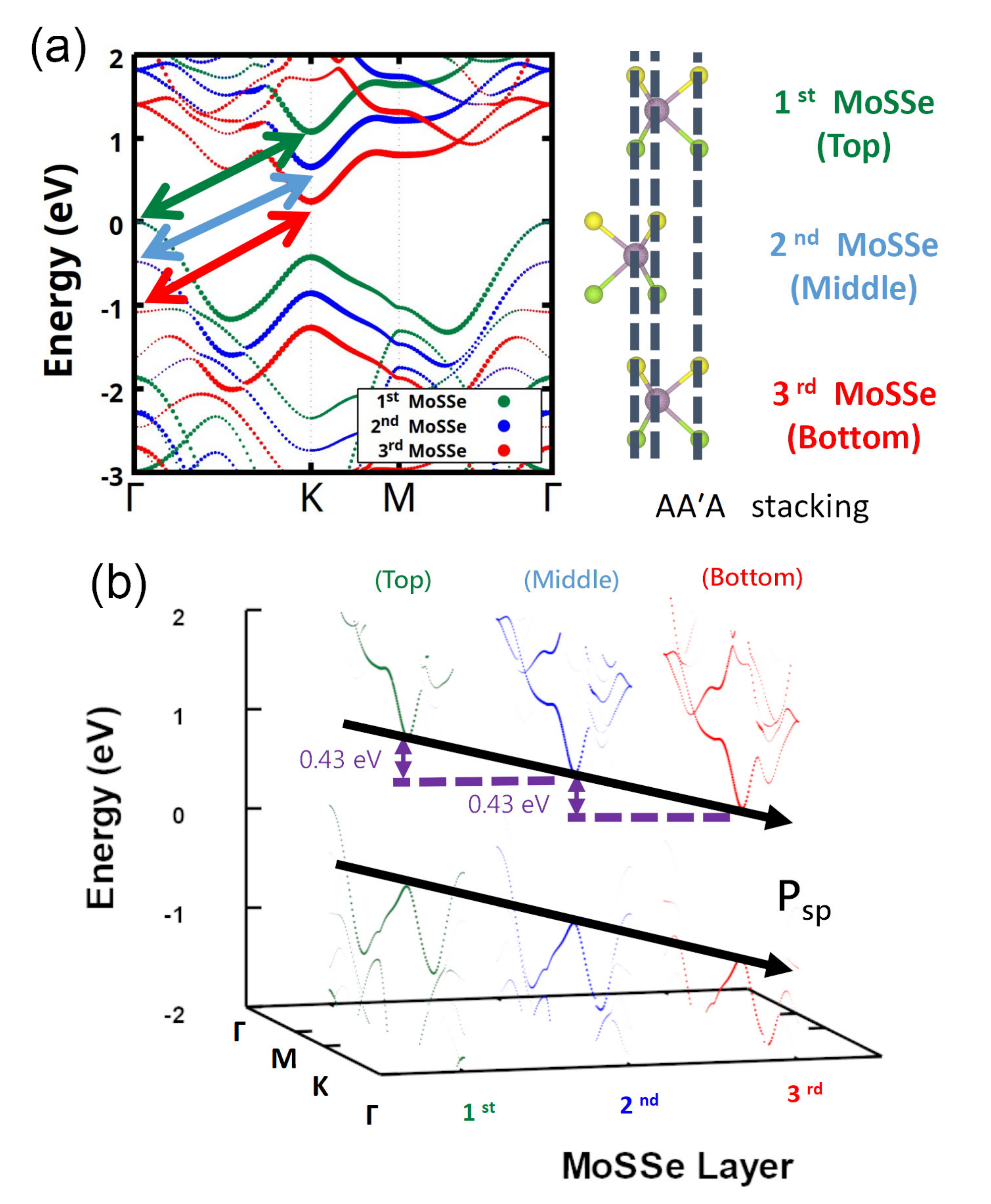}
\caption{(a) Projected band structure of AA$'$A-stacked MoSSe. The yellow, blue, and red dots indicate the top, middle, and bottom layers of MoSSe, respectively. (b) AA$'$A-stacked MoSSe projected band with layer separation in side view. Each layer offsets the other in the projected band diagrams of separate MoSSe layers, creating a spontaneous polarization (black arrow) through three layers.}
\label{MoSSe_Psp}
\end{figure}

DFT calculations are also used to obtain the device simulation parameters and the electronic parameters MoSSe, such as band gap, affinity, dielectric constant, and effective mass. Figure~\ref{MoSSe_Psp}(a) shows the band diagram of the AA$'$A MoSSe stacking with a relaxed lattice constant of 3.25~\AA\ and an interlayer distance of 3.19~\AA. All three layers exhibit an indirect band gap with a $K$ valley in the conduction bands and a $\Gamma$ valley in the valence bands. The projected band structure of each MoSSe layer can be observed in Fig.~\ref{MoSSe_Psp}(b). The yellow, blue, and red dots represent the projections of the top, middle, and bottom MoSSe layers, respectively, and are correlated with the colored arrows in Fig.~\ref{MoSSe_Psp}(a). As the layers' band components were separated, we observed that the bands were shifted by 0.43~eV as each MoSSe layer was added. This effect originates from the intrinsic dipole in the out-of-plane direction of MoSSe. Similar to the case of gallium nitride, we can represent this dipole effect by a spontaneous polarization $P_\mathrm{sp}$ term in the device simulations, which is illustrated by the two diagonal black arrows in  Fig.~\ref{MoSSe_Psp}(b). The $P_\mathrm{sp}$ term for MoSSe can be written as the product of the out-of-plane dielectric constant $\epsilon_{\bot}$ and the intrinsic electric field $E$:
 \begin{equation}
 P_\mathrm{sp}=\epsilon_{\bot} E.
     \label{psp}
 \end{equation}
It is easy to obtain $E$ from the thickness of the MoSSe and the value of the out-of-plane dipole. However, a problem arises when attempting to obtain $\epsilon_{\bot}$ in DFT calculations, because the vacuum layer along the $z$ axis will also be included. The principle of the equivalent capacitance method\cite{2D_dieletric_seperation} is adopted here to separate the dielectric constants of  MoSSe from those of the whole system. Carrier mobility is obtained using the Monte Carlo procedure described by Pai and Wu.\cite{Pai}

\begin{figure}[!t]
\includegraphics[width=\columnwidth]{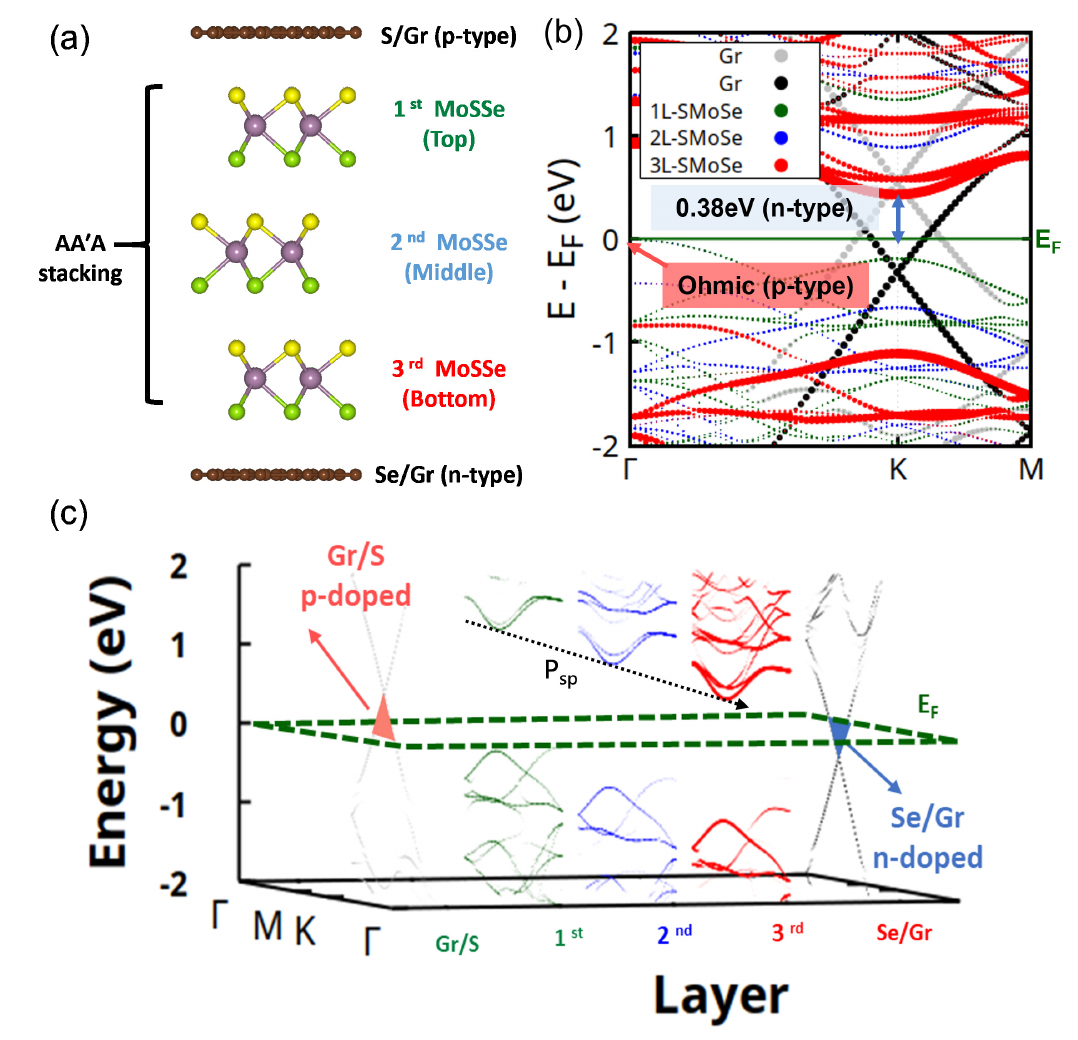}
\caption{(a) Illustration of calculated Gr/AA$'$A-MoSSe/Gr heterostructure system with the same colored-layer index as FIG. ~\ref{MoSSe_Psp}. The gray and black parts are the contributions Gr. (b) Projected band structures of the Gr/AA$'$A-MoSSe/Gr system with a 0.38 eV Shockey n-type contact and a p-type Ohmic contact. (c) The layer separation plot in the side view of the Gr/AA$'$A-MoSSe/Gr system. The spontaneous polarization caused intrinsic p-type and n-type doping on Gr at the S and Se side attachment, respectively.}
\label{Gr_psp}
\end{figure}

The aforementioned $P_\mathrm{sp}$ also influences the interactions between MoSSe and contacts. Initially, we construct the Gr/AA$'$A-MoSSe/Gr heterostructure, with the S and Se sides attached to the Gr electrodes, resulting in a lattice mismatch of 1.11\% between MoSSe and Gr. In Figure~\ref{Gr_psp}(b), the projected band structures of the Gr/AA$'$A-MoSSe/Gr system are illustrated. Notably, when Gr contacts a three-layer MoSSe material on the Se side (depicted in red), the Fermi level of the entire system is observed to be 0.38 eV lower than the conduction band of the MoSSe material, indicating an n-type Schottky barrier. Conversely, when Gr contacts a three-layer MoSSe material on the S side (depicted in green), a p-type Ohmic contact is observed at the $\Gamma$ point. These findings align with previous studies on JTMD contacts.\cite{JTMD_Graphene_pin, JTMD_Graphene_Hetero} According to our DFT calculation, monolayer MoSSe exhibits a strong $2.1\times10^6$ V/cm intrinsic dipole field in the out-of-plane direction. This characteristic may explain why this single material exhibits both n- and p-type contact formations when different sides of MoSSe attach to Gr. Furthermore, in Figure~\ref{Gr_psp}(c), the effect of the intrinsic dipole on the system is depicted; the $P_\mathrm{sp}$ pulls the Gr downward as the Se side contacts the Gr. Consequently, the Fermi level becomes higher than the Dirac cone, resulting in intrinsically n-doped Gr. Due to the tilting band edge caused by $P_\mathrm{sp}$, the energy difference between the conduction band edge and the Fermi level is smaller than the valence-band edge offset at the Se/Gr side, suggesting the formation of an n region. On the other hand, the S-side attachment elevates the band energy of the Gr, with the valence-band edge intersecting the Fermi level, indicating an intrinsic p-doped Gr and a p-contact region. Nevertheless, the Dirac cone of both Gr in the band structure retains its integrity, indicating the presence of physisorption characteristics between the two material systems.\cite{M_Gr}

\begin{figure}[!t]
\includegraphics[width=\columnwidth]{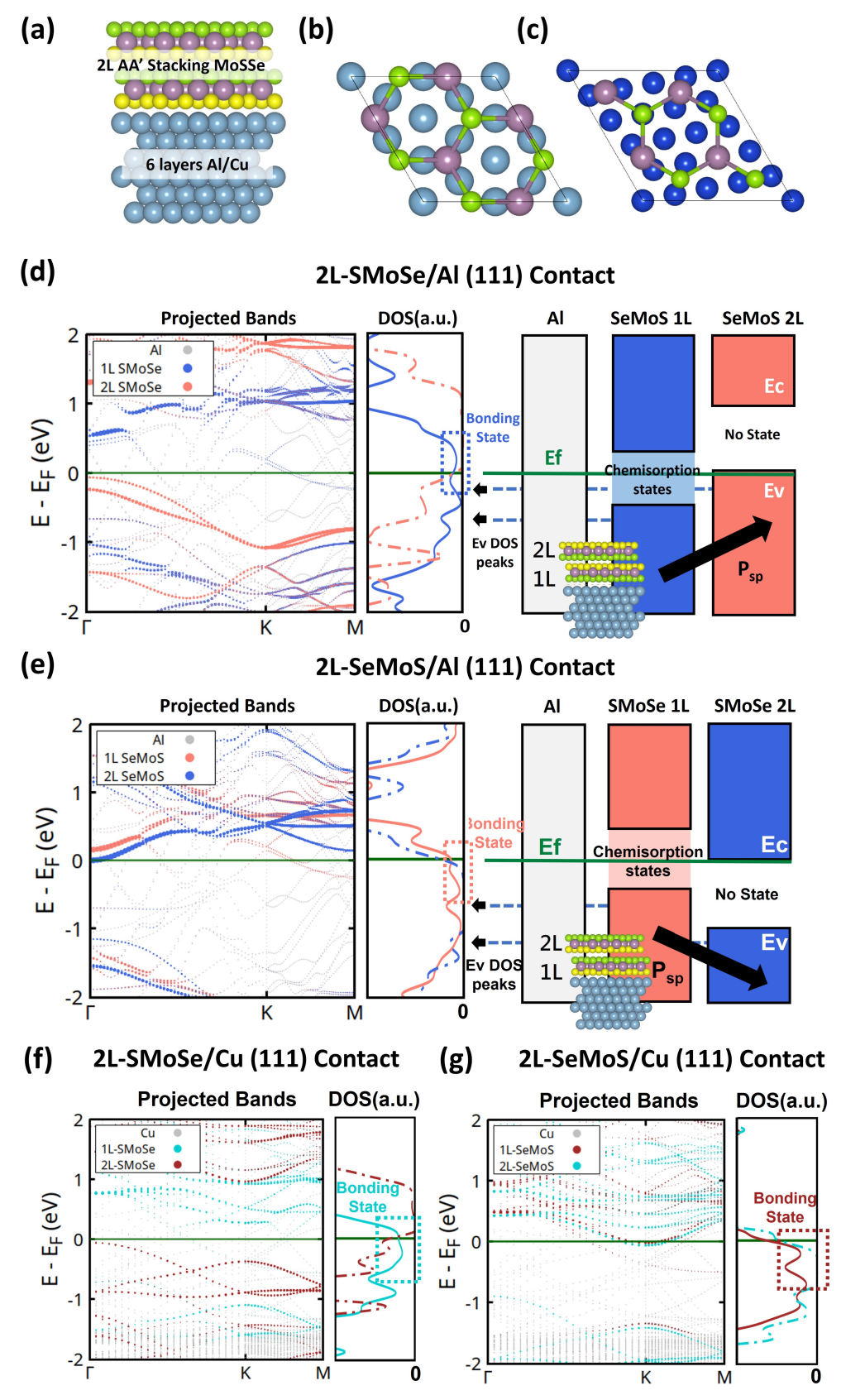}
\caption{(a) Illustration of calculated 2L-MoSSe/metal heterostructure system. (b) and (c) Top views of 2L-MoSSe/Al (111), and 2L-MoSSe/Cu (111) systems, respectively. (d) and (e) Projected band structures and 2L-MoSSe DOS of 2L-MoSSe/Al with Se-side and S-side contact respectively. The band-edge correspondence diagram is provided on the right side of the figure for reference, illustrating the relationship between the Fermi level (Ef) and the conduction band (Ec) and valence band (Ev).
(f) and (g) Projected band structures and 2L-MoSSe DOS of MoSSe Se-side and S-side contact with Cu, respectively. The gray parts are the contributions of metals.}
\label{M_Contact}
\end{figure}

We also simulate metal contact systems due to the interest in metal electrode cases, where common electrode materials such as Al and Cu are selected.\cite{TMD_contact_1, TMD_contact_2, TMD_contact_3, TMD_contact_4, TMD_contact_5} Fig.~\ref{M_Contact}(a) illustrates the simulated 2L-MoSSe/metal heterostructure and the top views are shown in Figs.~\ref{M_Contact}(b) and~\ref{M_Contact}(c), where the lattice mismatches are 0.44\%, and 2.97\% for the 2L-MoSSe/Al and 2L-MoSSe/Cu systems. Figures~\ref{M_Contact}(d)--\ref{M_Contact}(g) show the projected band structures, and Figs.~\ref{M_Contact}(d)--\ref{M_Contact}(f) represent the Se-face contact of MoSSe with the electrode material, and Figs.~\ref{M_Contact}(e)--\ref{M_Contact}(g) represent the S-face attachment. 

For Al (111) face attachment to the Se and S faces of bilayer MoSSe (2L-MoSSe) [Figs.~\ref{M_Contact}(d) and~\ref{M_Contact}(e)], the 2L-MoSSe band intersects the Fermi level, indicating that Ohmic contact occurs between the metal and 2L-MoSSe states. In contrast to the 2L-MoSSe/Gr cases, these metal contact projected band structures demonstrate the chemisorption characteristics due to significant hybridization in the band diagram. Different from typical TMDs, the internal electric field in JTMD plays a crucial role in augmenting the density of interface states. Specifically, in MoSSe, this electric field fosters the generation of electronic energy levels at the metal-semiconductor interface via chemisorptionlike bonding, leading to an increased density of interface states between the two materials\cite{JTMD_Metal_Hetero}. These additional bonding states, resulting from $P_\mathrm{sp}$ in MoSSe, are delineated in density of states (DOS) plots for all systems, indicated by a dotted frame.\cite{M_Gr} Due to the additional bonding states caused by chemical adsorption, it becomes challenging to correlate the band edge of MoSSe closest to the metal layer. 

We correlate the band-edge distribution of the closest-to-metal layer of MoSSe (1L MoSSe) based on the direction of $P_{sp}$ previously mentioned and the peak present at the Ev edge of MoSSe layers, and depict the corresponding band edge with DOS on the right side of Figs.~\ref{M_Contact}(d) and (e). When the Se face contacts with Al [Fig.~\ref{M_Contact}(d)], it can be observed that the Fermi level coincides with the outer MoSSe (2L MoSSe) at the valence-band edge. Due to the upward direction of $P_{sp}$, the energy band of the inner layer of MoSSe (1L MoSSe) shifts towards lower energy. Supported by the peak at the Ev edge, it can be inferred that the position of the Fermi level is closer to the conduction band, leaning towards an n-type contact. Conversely, when the S face contacts with Al [Fig.~\ref{M_Contact}(e)], with the downward direction of $P_{sp}$, the energy band of the 1L MoSSe shifts towards higher energy. A comparison reveals that the Fermi level is situated in the middle of the band, leaning towards a p-type contact. Although we correlate the n- or p-type contact in both systems, there are numerous bonding states present at the gap position of 1L MoSSe, indicating the Ohmic contact formation. In other words, both electrons and holes can freely flow through this low-tunneling barrier Ohmic contact, thereby enhancing device performance.\cite{M_Gr, JTMD_Metal_Hetero} Similar n- and p-type Ohmic contact formation results for different face attachments are also seen in the case of Cu (111) [Fig.~\ref{M_Contact}(f) and (g)], indicating Cu can also bring Ohmic contact properties as 2L-MoSSe attaches. The chemisorptionlike property imparts the advantage concerning tunneling barrier reduction at the contact region, thus improving device performance. \cite{JTMD_Metal_Hetero, M_Gr}

\begin{figure}
\resizebox{75mm}{!}{\includegraphics{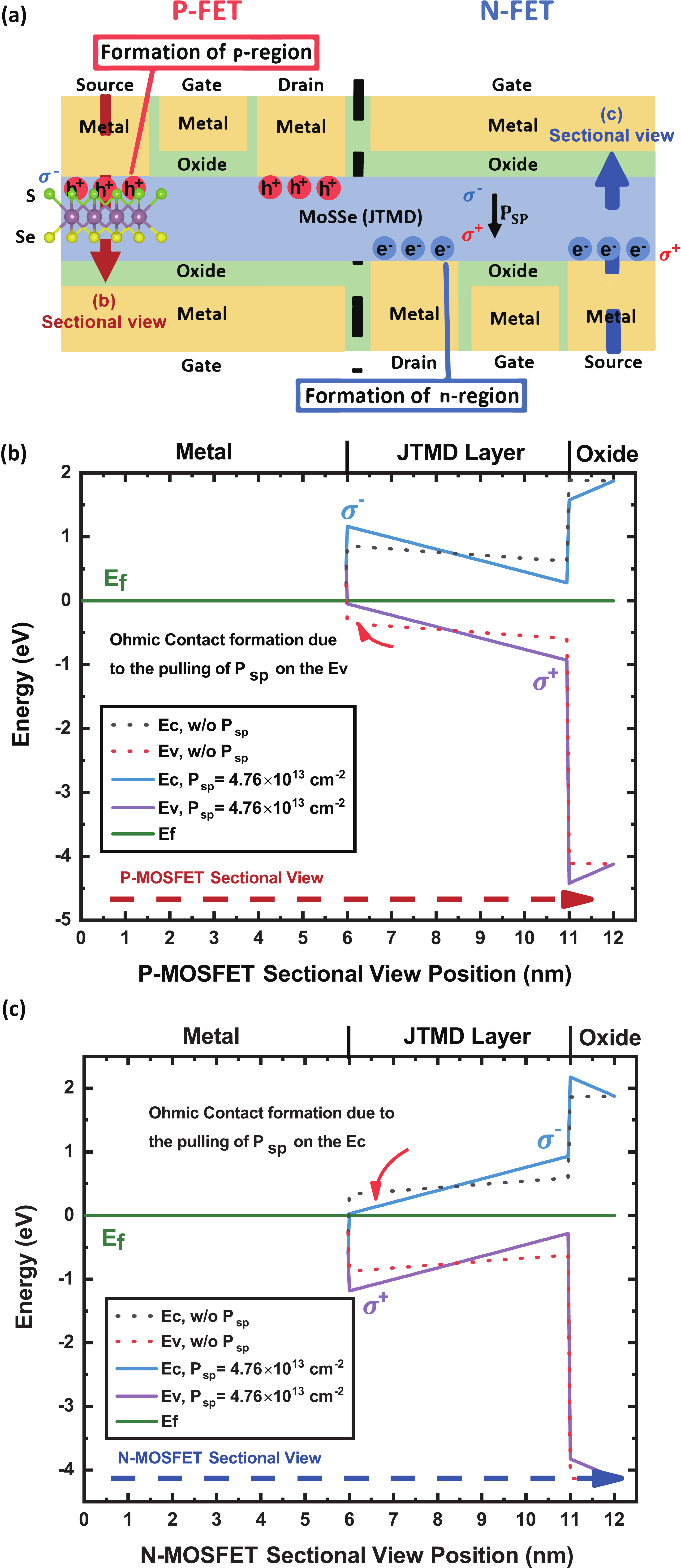}}
\caption{(a) Schematic of 2D MoSSe JTMD CMOSFET device. (b) and (c) Energy band and carrier concentration sectional-view diagrams of p-MOSFET and n-MOSFET, respectively, from source metal to gate oxide. (d) n-MOSFET sectional-view conduction-band diagram and electron density with different $P_\mathrm{sp}$ values for comparison.}
\label{Janus_CMOSFET}
\end{figure}

Figure~\ref{Janus_CMOSFET}(a) is a schematic of double-gate CMOSFET device design exploiting the strong $P_\mathrm{sp}$ properties of MoSSe. As illustrated to the left of the black dashed line, to create a p-MOSFET,  two metal electrodes are attached to the S face to produce a p-type source and drain regions, while the remaining MoSSe region between them acts like a channel area after the oxide and gate metal have been placed in the vertical direction. An n-MOSFET can be created similarly, as illustrated to the right of the dashed line, with the only difference being that in this case, it is the Se face, rather than the S face, to which metals are attached from n-type regions at the source and drain. This horizontal CMOSFET design has two advantages: (1) the Janus MoSSe reduces the complexity of material selection since only one channel material is used to form n- and p-MOSFETs; (2) this structure enables layer-by-layer buildup during the fabrication process, which makes the process more reliable. 

To investigate the electron-band characteristics of the designed MoSSe MOSFETs, the 2D DDCC solver is used to simulate device characteristics. We design our device using MoSSe of 5-nm channel thickness to represent MoSSe multilayers since this thickness ensures there is sufficient charge flowing in the channel region. All the parameters used in the 2D DDCC simulation are given in Table~\ref{parameters}. In our device simulation setting, the metal work function is set at the midgap of the MoSSe to showcase the possibility of n- and p-type contact formation even in the worst-case metal selection scenario. Cross-section views of the band diagram are shown in Figs.~\ref{Janus_CMOSFET}(b) and~\ref{Janus_CMOSFET}(c) for the p- and n-MOSFETs, respectively. The strong $P_\mathrm{sp}$ in the MoSSe vertical direction acts like partial positive and negative charges attracting electrons and holes. The negative polarization charge causes the Ec and Ev to be pulled upward [Fig.~\ref{Janus_CMOSFET}(b)]  and negative polarization charge casues the Ec and Ev band to be bent downward [Fig.~\ref{Janus_CMOSFET}(c)] at the interface of the MoSSe region. This band-edge bending, originating from $P_\mathrm{sp}$, plays a pivotal role in facilitating the formation of Ohmic contacts within the device, which is in good agreement with the aforementioned DFT calculations. The difference in physical behavior between the different sides of MoSSe attachment corresponds simply to the difference in sign of $P_\mathrm{sp}$ here. 
In contrast, when simulating TMDs lacking this intrinsic dipole, where $P_\mathrm{sp}$ is set to zero (dot lines labeled "w/o $P_\mathrm{sp}$"), the same device structure is employed. In the absence of $P_\mathrm{sp}$ assistance, the formation of Ohmic contacts falls short, resulting in the Schottky barrier at the interface.

\begin{figure}[!t]
\includegraphics[width=\columnwidth]{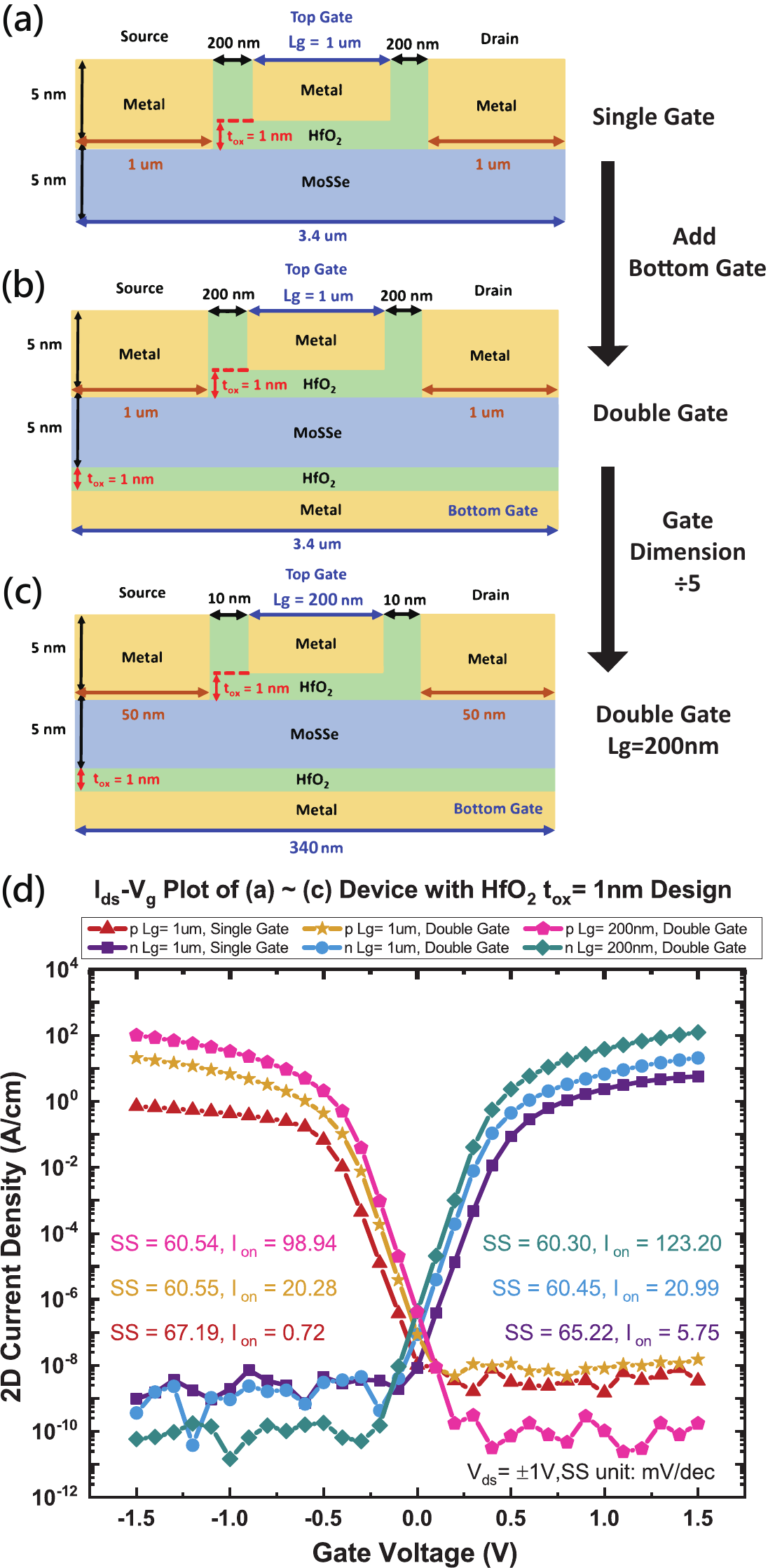}
\caption{(a) and (b) Schematics of single- and double-gate MoSSe MOSFET designs, respectively, with $L_g= 1~\mu$m, showing the structure and critical device dimensions. (c) Schematic of double-gate MoSSe MOSFET design with $L_g= 200$~nm. (d) $I_{ds}$-$V_{g}$ plots for the designs shown in (a)--(c) with a $t_\mathrm{ox}= 1$~nm HfO$_2$ gate oxide for both polarities of the MOSFET.  Parameters are given in Table~\ref{parameters}.}
\label{large_small_FET}
\end{figure}

The $I$-$V$ characteristics of several MoSSe MOSFET designs are now investigated. Figure~\ref{large_small_FET}(a) shows a single-gate design with gate length $L_g=1~\mu$m, which is a common dimension adopted in experimental studies of FETs based on 2D  materials\cite{CFET_IEEE,2D_FET_Bench}. Device modifications such as the use of double gates [Fig.~\ref{large_small_FET}(b)] and smaller dimension [Fig.~\ref{large_small_FET}(c)]  are also simulated. The simulations for all devices assume an HfO$_2$ gate oxide of thickness $t_\mathrm{ox}=1$~nm. Figure~\ref{large_small_FET}(d) shows the transfer characteristics ($I_{ds}$-$V_{g}$) of the n- and p-MOSFETs from  Figs.~\ref{large_small_FET}(a)--\ref{large_small_FET}(c). The curves with squares, circles, and diamonds correspond to the n-MOSFETs in (a), (b), and (c), respectively, and those with triangles, stars, and pentagons to the p-MOSFETs in (a), (b), and (c), respectively. The key distinction lies in the opposite orientation of $P_\mathrm{sp}$ in the JTMD layer. The typical $I_{ds}$-$V_{g}$ curves for the n- and p-MOSFETs indicate that all three devices exhibit satisfactory enhancement-mode MOSFET behavior. Among the three devices, the single-gate design [Figure~\ref{large_small_FET}(a)] has the most significant subthreshold swing (SS) and the lowest \footnotesize ON\normalsize current $I_\mathrm{ON}$. When a bottom gate is added [Figure~\ref{large_small_FET}(b)], the SS drops to around 60.5~mV/dec, while $I_\mathrm{ON}$ increases by an order of magnitude, indicating an improvement in gate control. Having shown that both designs are feasible at a larger scale, we reduce the horizontal scale of the double-gate design to $L_g= 200$~nm  [Figure~\ref{large_small_FET}(c)] to examine the device performance at a smaller scale. The proximity of the SS values in the n-MOSFET suggests that reducing $L_g$ to 200~nm would not significantly impact the control of the gate in this design. In addition, $I_\mathrm{ON}$ is increased by an order of magnitude compared with that of the larger-scale device, while the 
\footnotesize OFF\normalsize-state current $I_\mathrm{OFF}$ is decreased by about an order, indicating an improvement in $I_\mathrm{ON}/I_\mathrm{OFF}$. All of these results show that the SS and \footnotesize ON- \normalsize or \footnotesize OFF\normalsize-current characteristics of the designs considered are those of a typical MOSFET device. The lower SS and higher $I_\mathrm{ON}$ consistently observed for all the n-MOSFETs compared with the corresponding p-MOSFETs may originate from the slightly higher mobility of electrons compared with holes.

\begin{figure}[!t]
\includegraphics[width=0.9\columnwidth]{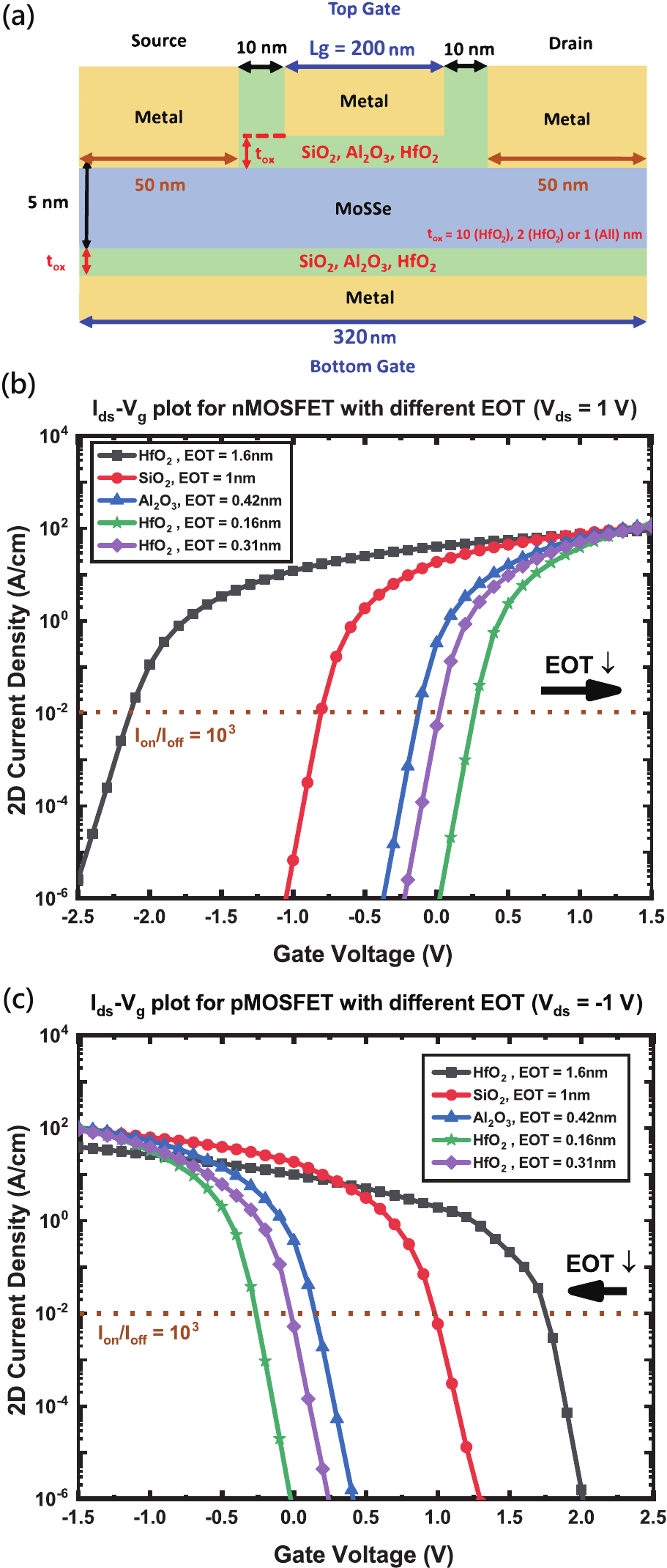}
\caption{(a) Schematic of double-gate MoSSe MOSFET with $L_g= 200$~nm  and a variety of oxides with different values of EOT. (b) and (c) $I_{ds}$-$V_{g}$ plots for n- and p-MOSFETs, respectively, with different values of EOT. $t_\mathrm{ox}= 1$~nm  for all oxides, and, in addition, results for designs with $t_\mathrm{ox}= 2$ and 10~nm  are also shown for HfO$_2$. The parameters are given in Table~\ref{parameters}.}
\label{np_FET}
\end{figure}

For the successful operation of a CMOSFET,  an enhancement-mode MOSFET device is essential. Since the gate oxide thickness $t_\mathrm{ox}$ plays a crucial role in determining the operation mode of a MOSFET, a variety of oxides and different values of $t_\mathrm{ox}$, are tested to determine the optimal design of the device in Fig.~\ref{np_FET}(a)  in terms of the equivalent oxide thickness (EOT), and the results are shown in Figs.~\ref{np_FET}(b) and~\ref{np_FET}(c) for the n- and p-MOSFETs, respectively. Here, we define \footnotesize ON \normalsize or \footnotesize OFF\normalsize state switching when $I_\mathrm{ON}/I_\mathrm{OFF}>10^3$. First, a SiO$_2$ oxide gate of thickness 1~nm  is tested, and the $I_d$-$V_g$ result is shown by the curve with red circles. The device is clearly in the \footnotesize ON \normalsize state ($I_d \sim 90$~A/cm) when no gate bias is applied, indicating that the MOSFET is a depletion-mode device. To search for the threshold  EOT at which the device switches to the enhancement mode, we try decreasing the EOT by selecting an Al$_2$O$_3$ gate with $t_\mathrm{ox}1$~nm (EOT $=$ 0.42~nm, blue triangles). However, the current remains of the order of $10^1$~A/cm at $V_g= 0$~V, and both polarities remain in the depletion mode. Therefore, a high-$\kappa$ material, namely, HfO$_2$, is selected to help decrease the EOT of the MOSFETs. Using HfO$_2$ with $t_\mathrm{ox}=1$~nm (EOT $=$ 0.16~nm, green stars) gives a superior enhancement-mode device for both n- and p-MOSFETs. Finally, we optimize the $t_\mathrm{ox}$ of the HfO$_2$  to 2~nm (EOT $=$ 0.31~nm, purple diamonds) to prevent current leakage through the thin gate oxide. As can be seen from Figs.~\ref{np_FET}(b) and~\ref{np_FET}(c), both of the $I_d$--$V_g$ curves for this optimized value of $t_\mathrm{ox}$ intersect the \footnotesize OFF \normalsize state brown dashed line at $V_g=0$~V, indicating preservation of the enhancement mode even though the EOT has increased. These results for HfO$_2$ with $t_\mathrm{ox}= 2$~nm suggest that the EOT threshold between the depletion and enhancement modes in this MOSFET design is 0.312~nm.  For this optimal  MOSFET design based on HfO$_2$ with $t_\mathrm{ox}=2$~nm,  the SS as extracted from Figs.~\ref{np_FET}(b) and~\ref{np_FET}(c) is about 63~mV/dec.  
Since such low-$t_\mathrm{ox}$ devices may encounter several processing difficulties, simulation results for a more easily fabricated thick HfO$_2$ gate with $t_\mathrm{ox}= 10$~nm (EOT $=$ 1.6 nm, black squares)  are also given in Figs.~\ref{np_FET}(b) and~\ref{np_FET}(c). Owing to the increased EOT, both the n- and p-type MOSFETs exhibit the depletion mode in terms of transfer characteristics, where both the $I_\mathrm{ON}$ drops and the device turns off at $-$2.1~V and 1.8~V for n- and p-MOSFETs, respectively. All in all, the $I$-$V$ plots demonstrate the possibility of creating a workable MOSFET even with larger values of $t_\mathrm{ox}$ and provide a reference for future experimental studies.

\begin{figure}[!t]
\includegraphics[width=0.9\columnwidth]{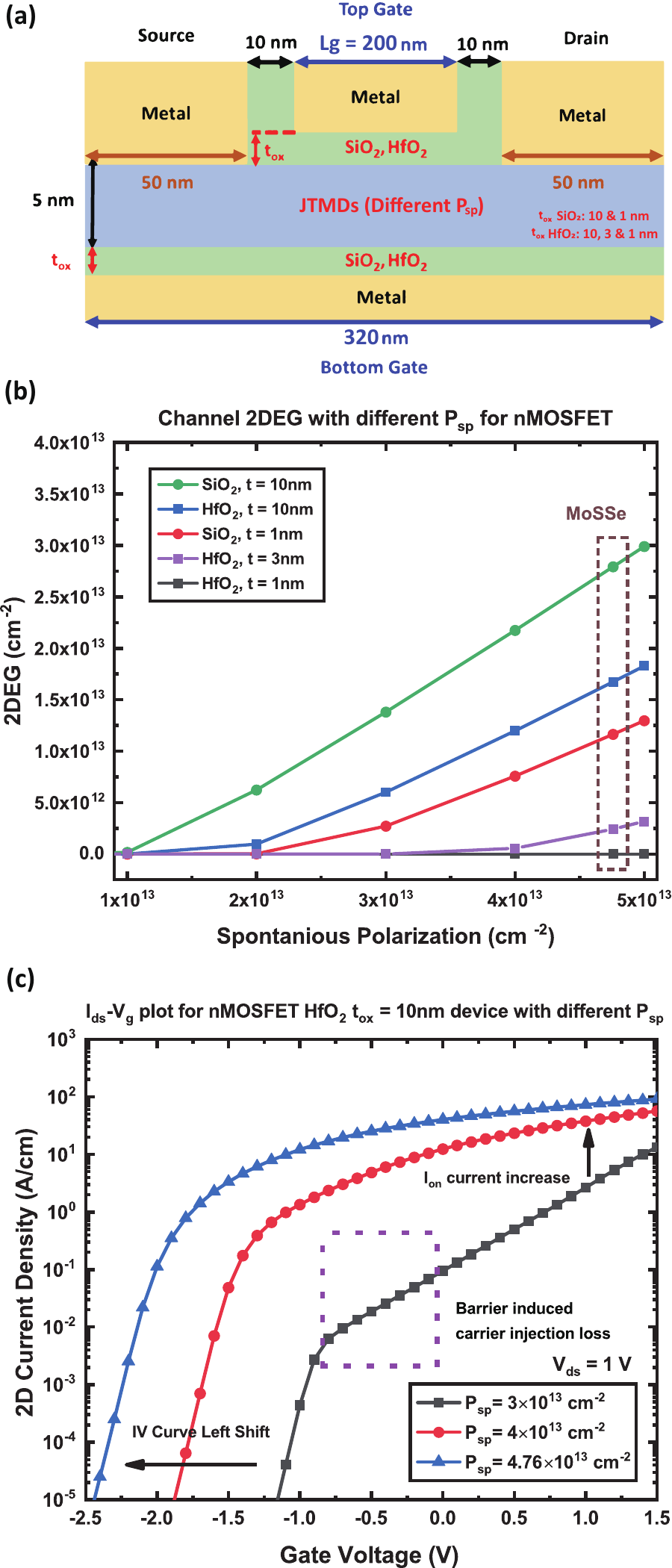}
\caption{(a) Schematic of $L_g= 200$~nm double-gate MoSSe MOSFET with selected oxides and JTMDs with different values of EOT and $P_\mathrm{sp}$, respectively. (b) Channel 2DEG versus different $P_\mathrm{sp}$ magnitude plots for n-MOSFET in linear scale. (c) $I_{ds}$--$V_{g}$ plots for n-MOSFET HfO$_2$ $t_\mathrm{ox}= 10$~nm device, with different $P_\mathrm{sp}$ for comparison.}
\label{Psp_FET_comparison}
\end{figure}

It is also worth considering how the value of $P_\mathrm{sp}$ affects this carrier accumulation phenomenon at the channel MoSSe region. To reveal the impact of the intrinsic dipole's magnitude on device performance, we calculate the two-dimensional electron gas (2DEG) in the n-MOSFET device with a 200-nm length channel, shown in [Figure~\ref{Psp_FET_comparison}(a)], with various assumed $P_\mathrm{sp}$ values ranging from $1\times10^{13}$ to $5\times10^{13}$ ~cm$^{-2}$. These $P_\mathrm{sp}$ values can be regarded as JTMDs with different intrinsic dipole strengths, providing insights into how the $P_\mathrm{sp}$ magnitude influences the device's 2DEG. The 2DEG corresponding to MoSSe’s $P_\mathrm{sp}$ is specified in the black dot frame. SiO$_2$ ($t_\mathrm{ox}= 1$ and $10$~nm) and HfO$_2$ ($t_\mathrm{ox}= 1, 3$ and $10$~nm) were selected as representative low-$\kappa$ and high-$\kappa$ gate oxides for this study. To isolate the impact of $P_\mathrm{sp}$, the calculations were performed at zero bias, corresponding to the equilibrium state.

Figure~\ref{Psp_FET_comparison}(b) illustrates the influence of $P_\mathrm{sp}$ on the 2DEG concentration. In cases with high EOT (SiO$_2$, $t_\mathrm{ox}= 10$~nm), 2DEG exhibits a nearly linear increase with increasing $P_\mathrm{sp}$. This linearity arises because the high EOT allows carriers to reach saturation within the investigated $P_\mathrm{sp}$ range ($1\times10^{13}$ to $5\times10^{13}$~cm$^{-2}$). However, as EOT decreases, the starting point of the linear region shifts towards higher $P_\mathrm{sp}$ values. For instance, the linear region begins at $2\times10^{13}$~cm$^{-2}$ for HfO$_2$ with $t_\mathrm{ox}= 10$~nm and $3\times10^{13}$~cm$^{-2}$ for SiO$_2$ with $t_\mathrm{ox}= 1$~nm. The $P_\mathrm{sp}$ value at the onset of the linear regime serves as an indicator for achieving enhanced mode devices. The HfO$_2$ design with $t_\mathrm{ox}= 1$~nm has the lowest EOT, resulting in a linear regime starting beyond $5\times10^{13}$~cm$^{-2}$. This suggests that MoSSe (marked by the brown dotted box) will operate in enhanced mode due to the initially low 2DEG concentration. Conversely, increasing the HfO$_2$ thickness to 3~nm lowers the EOT and brings the linear region starting point to $4\times10^{13}$ ~cm$^{-2}$. This indicates a depletion mode behavior for MoSSe with HfO$_2$ ($t_\mathrm{ox}= 3$~nm) due to the higher initial induced 2DEG.

The influence of $P_\mathrm{sp}$ extends to the \textit{I-V} characteristics of MOSFETs. To focus on its impact on the linear 2DEG region, we examined devices with $P_\mathrm{sp}$ values of \(3\times10^{13}\), \(4\times10^{13}\), and \(4.76\times10^{13}\)~cm\(^{-2}\), all designed with a 10 nm thick HfO\(_2\). As shown in Figure~\ref{Psp_FET_comparison}(c), an increased $P_\mathrm{sp}$ shifts the transfer \textit{I-V} curve towards more negative voltages. This shift reflects the need for a stronger gate voltage to deplete the augmented 2DEG in the channel. Moreover, the higher carrier concentration resulting from a higher $P_\mathrm{sp}$ leads to a greater \footnotesize ON \normalsize current (I\(_{ON}\)). 
Additionally, $P_\mathrm{sp}$ can influence carrier injection at the contacts. For strengths of \(4.76\times10^{13}\) and \(4\times10^{13}\)~cm\(^{-2}\) value of $P_\mathrm{sp}$, the device also has a larger g$_m$ value due to a smaller contact resistance. However, if $P_\mathrm{sp}$ falls below a certain threshold, devices display nonideal \textit{I-V} curves near saturation (highlighted by the purple dotted box). As detailed in Figure~\ref{Janus_CMOSFET}(c), insufficient $P_\mathrm{sp}$ compromises Ohmic contacts, resulting in increased resistance from an additional barrier at the contact regions. This issue can be mitigated by selecting a metal with a work function closer to the band edge to offset the weaker band bending at the contact interface. For p-MOSFETs, the same principles apply.

\begin{figure}[!t]
\includegraphics[width=\columnwidth]{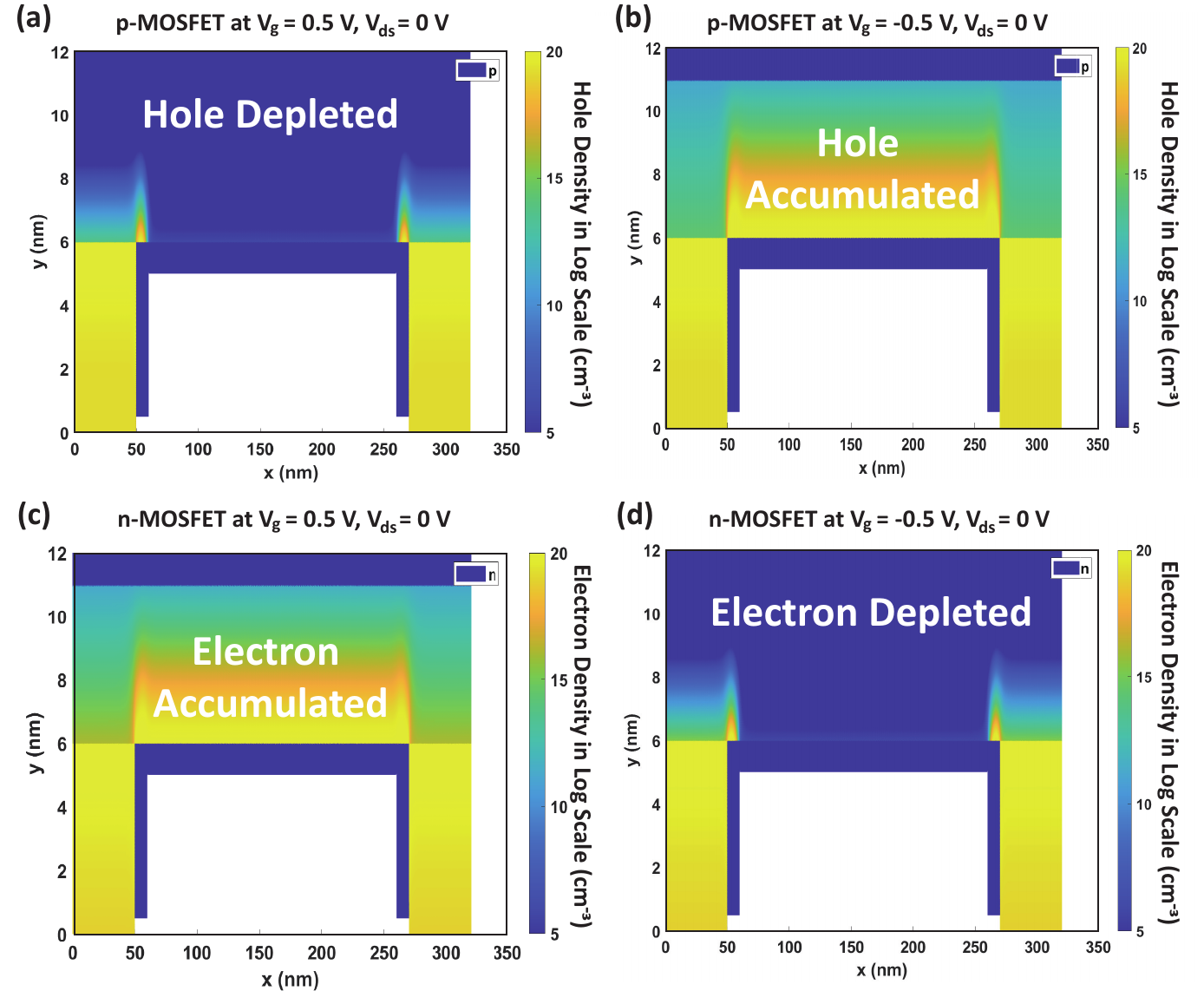}
\caption{Simulated majority carrier concentration plots for double-gate HfO$_2$ (a) p-MOSFET under $V_g= -0.5$~V, (b)  p-MOSFET under $V_g= 0.5$~V, (c) n-MOSFET under $V_g= -0.5$~V, and (d) n-MOSFET under $V_g= -0.5$~V bias operation. For observational convenience, the $V_{ds}$ bias is set to zero.}
\label{Carrier_Dis_ONOFF}
\end{figure}

To understand the distribution of carriers in the device, we export the two-dimensional distribution map of the majority carrier of the device under different operating conditions in Figure ~\ref{Carrier_Dis_ONOFF}, which demonstrates the \footnotesize ON\normalsize or\footnotesize OFF \normalsize characteristics of the p- and n-MOSFET devices with HfO$_{2}$ gate with $t_\mathrm{ox}= 1$~nm. The source and drain areas are defined as those where the MoSSe is attached to the metal, which gives a relatively high carrier concentration of about $10^{14}$~cm$^{-3}$ (green color). The carrier concentration gradually decays as the position moves away from the contact region. Owing to the 10-nm gate metal isolation design, two bumplike regions of high carrier concentration appear next to the source and drain areas. The thicker oxide eliminates the dipole effect of MoSSe, and thus carrier gas remains. Taking the p-MOSFET as an example, we can see that with the application of a negative $-$0.5 V gate bias in Fig. ~\ref{Carrier_Dis_ONOFF}(a), holes are attracted into the channel region, forming a strong inversion channel layer and causing the device to turn on. By contrast, when a positive 0.5~V gate bias is applied in Fig. ~\ref{Carrier_Dis_ONOFF}(b), holes are depleted, and the inversion layer disappears, indicating an \footnotesize OFF\normalsize-state device. A similar result is also obtained for the n-MOSFET, with the only difference being that a strong inversion layer appears when $V_g= 0.5$~V (Fig. ~\ref{Carrier_Dis_ONOFF}(c)) and the inversion layer vanishes when a $V_g= -0.5$~V bias is applied (Fig. ~\ref{Carrier_Dis_ONOFF}(d)).

\begin{figure}[!t]
\includegraphics[width=\columnwidth]{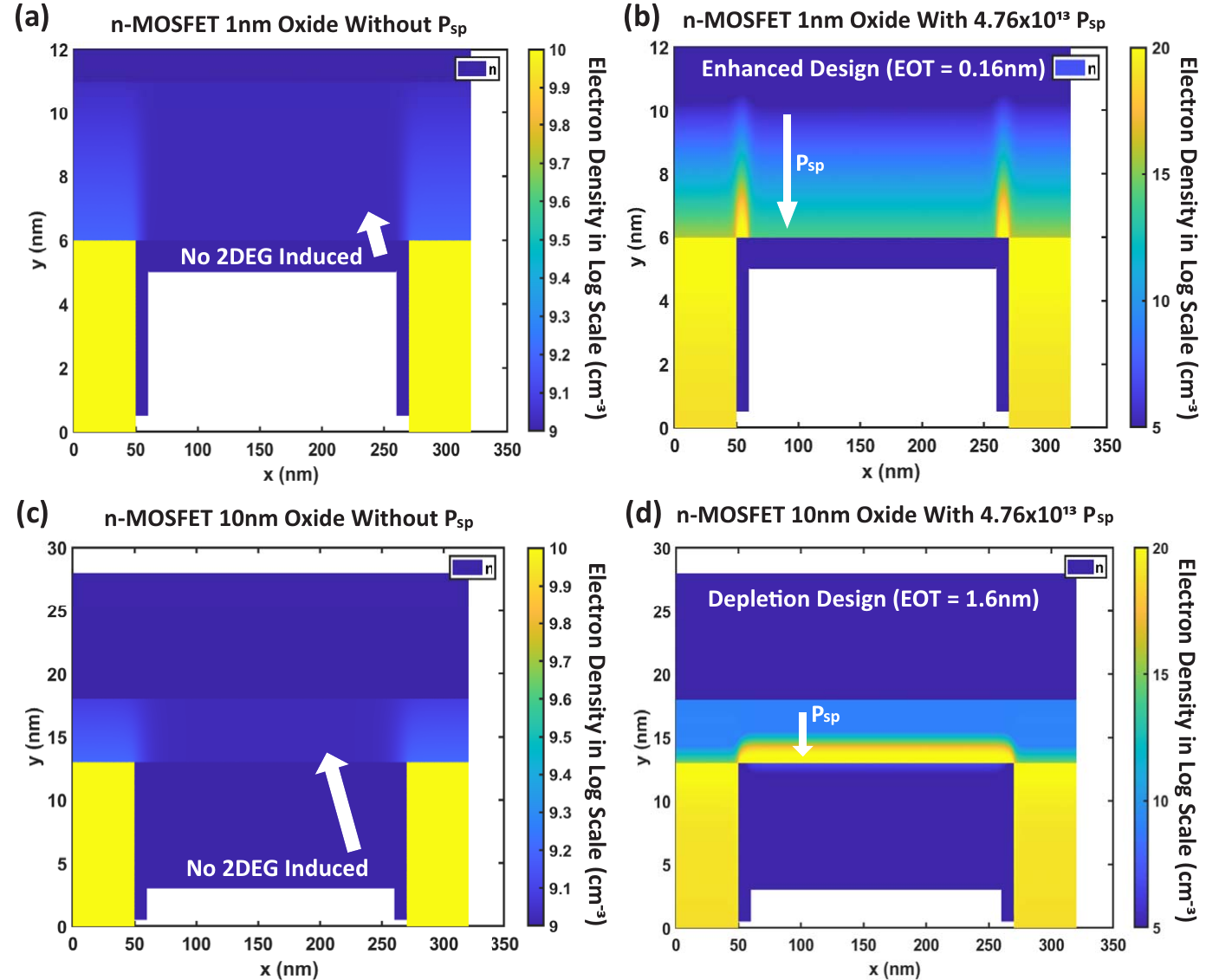}
\caption{n-MOSFET 2D majority carrier concentration plots of enhance (HfO$_2$ $t_\mathrm{ox}= 1$~nm) mode design (a) without $P_\mathrm{sp}$ and (b)  $4.76\times10^{13}$~cm$^{-2}$ $P_\mathrm{sp}$ in the channel layer. And n-MOSFET depletion (HfO$_2$ $t_\mathrm{ox}= 10$~nm) mode design (c) without $P_\mathrm{sp}$ and (d)  $4.76\times10^{13}$~cm$^{-2}$ $P_\mathrm{sp}$ in the channel layer. All results are under an equilibrium state to observe the effect of the intrinsic dipole on carrier distribution in the channel.}
\label{Carrier_Dis_Psp}
\end{figure}

Figure \ref{Carrier_Dis_Psp} demonstrates the influence of the intrinsic dipole on the n-MOSFET channel region using a 2D simulation of a double-gate n-MOSFET device operating in both depletion mode (HfO$_2$, $t_\mathrm{ox}= 10$ nm) and enhancement mode (HfO$_2$, $t_\mathrm{ox}= 1$ nm). An additional dataset labeled "Without $P_\mathrm{sp}$" is provided, representing traditional TMD materials, to highlight the impact of $P_\mathrm{sp}$ on carrier distribution in the MoSSe channel. All biases are zeroed to observe carrier distribution under equilibrium conditions. 
In configurations lacking an intrinsic dipole (as shown in Figs. ~\ref{Carrier_Dis_Psp}(a) and (c)), no 2DEG is observed across the channel region, with the electron concentration remaining around \(10^{10}\) ~cm\(^{-3}\) regardless of oxide thickness. In contrast, when an intrinsic dipole is present in the channel region, it acts as a built-in electric field within the channel layer. This field, influenced by $P_\mathrm{sp}$, is depicted in Figs. \ref{Carrier_Dis_Psp}(b) and (d). As a result, electrons align with the direction of $P_\mathrm{sp}$, forming a 2DEG akin to a conventional GaN-based n-FET channel. Any JTMD with a pronounced $P_\mathrm{sp}$ will display this characteristic 2DEG formation. The intensity of the 2DEG is mainly dictated by the EOT. A value of EOT= 0.16~nm corresponds to the enhancement mode (as seen in Fig.~\ref{Carrier_Dis_Psp}(b)). As EOT expands to 1.6~nm (illustrated in Fig.~\ref{Carrier_Dis_Psp}(d)), the electron concentration at the channel interface experiences a sharp increase, causing a shift from enhancement mode to depletion mode. In the case of p-MOSFETs, the majority carrier distribution mirrors that of n-MOSFETs, yielding analogous outcomes.

\section{Conclusions}
We have introduced an alternative CMOSFET design based on the intrinsic dipole properties of JTMD, with a specific focus on the material MoSSe. Our investigation has encompassed diverse heterostructures, including Gr/AA$'$A-MoSSe/Gr and 2L-MoSSe/Al (111) face and 2L-MoSSe/Cu (111) face structures, which we have analyzed using first-principles DFT. We have also examined the interactions between MoSSe, Gr, and Al and Cu electrodes, identifying characteristics, such as Schottky barriers and Ohmic contacts. Our design leverages the strong dipole field of MoSSe to create n- and p-type contact regions, thereby forming a CMOSFET with simplified material selection. The proposed horizontal CMOSFET architecture offers advantages in fabrication reliability and integration. 
Device simulations have provided insights into the spontaneous polarization caused by band bending and the related carrier distribution, with an optimized EOT $=$ 0.312~nm giving a favorable threshold between the depletion and enhancement modes of the MOSFET device. The discussion also delves into the influence of spontaneous polarization magnitude, a crucial factor determining the conditions for achieving Ohmic contact and the formation of the channel 2DEG and 2DHG. The majority of carrier plots show the capability of the device to control carriers in the MoSSe channel region, demonstrating its potential for low-voltage logic elements in integrated circuits.
On the basis of the surface polarization strategy adopted here, it might be possible to form a MOSFET from any JTMD, provided that the 2D material possesses a strong intrinsic out-of-plane dipole. Regardless, this work has introduced a promising approach to CMOSFET design, offering valuable insights into materials, interfaces, and device characteristics for future electronic applications.

\begin{acknowledgments}
This work was supported by the National Science and Technology Council under Grants No.~NSTC 112-2221-E-002-215-MY3, No.~NSTC 112-2923-E-002-002, and No.~112-2221-E-002-214-MY3.

Yun-Pin Chiu: conceptualization (equal); data curation (lead); formal analysis (equal); methodology (equal); validation (lead); visualization(lead); writing – original draft (lead). 
Hsin-Wen Huang: conceptualization (equal); data curation (support); formal analysis (equal); methodology (equal); validation (support). 
Yuh-Renn Wu conceptualization (equal); formal analysis (equal); funding acquisition (lead); project administration (lead); software(lead); supervision (lead); writing-–review and editing (lead).
The authors have no conflicts to disclose

\end{acknowledgments}

\section*{References}

\nocite{*}
\bibliography{Paper125055_refs_EDITED}

\end{document}